\documentclass{moriond}

\def\Journal#1#2#3#4{{#1} {\bf #2}, #3 (#4)}

\def\NPB{{\em Nucl. Phys.} B}

\def\PRD{{\em Phys. Rev.} D}

\def\CPC{{\em Chin.Phys.} C}

\def\EPJC{{\em Euro.Phys. Jour}. C}
\def\be{\begin{equation}}
\def\ee{\end{equation}}
\def\bea{\begin{eqnarray}}
\def\eea{\end{eqnarray}}

\begin{document}
\vspace*{4cm}
\title{CHARMLESS TWO-BODY   $B$ MESON DECAYS IN FACTORIZATION ASSISTED TOPOLOGICAL AMPLITUDE APPROACH}

\author{ Cai-Dian L\"u and Si-Hong Zhou  }

\address{Institute of High Energy Physics, P.O. Box 918, Beijing 100049,   China;\\
         %2.~State Key Laboratory of Theoretical Physics, Institute of Theoretical Physics,\\
         School of Physics, University of Chinese Academy of Sciences, Beijing 100049,  China
         }

\maketitle
\abstracts{
We analyze charmless two-body non-leptonic $B$ decays   under the framework
of factorization assisted topological amplitude approach. Unlike  the conventional flavor diagram approach,
we consider flavor $SU(3)$ breaking effect assisted by factorization hypothesis
for topological diagram amplitudes of different decay modes, by factorizing out the
corresponding decay constants and form factors.
The non-perturbative parameters of topology diagram magnitudes $\chi$ and strong phase $\phi$
are universal that can be extracted by $\chi^2$ fit from
current abundant experimental data of charmless $B$ decays. The number of free parameters and
the $\chi^2$ per degree of freedom are both reduced comparing with previous analysis. With these
best fitted parameters, we predict branching fractions and $CP$ asymmetry parameters of
nearly 100 $B_{u,d}$ and $B_s$ decay modes. The long-standing
$\pi \pi$ and $\pi K$-$CP$ puzzles are solved simultaneously.}

\section{Introduction}\label{sec:1}

Charmless two-body non-leptonic $B$ decays are of importance for testing the standard model(SM).
They can be used to study ${\it CP}$ violation via the interference of tree and penguin contributions.
They are also sensitive to signals of new physics that would change the small loop effects from
penguin diagrams. 
With regards to them, the BaBar,  Belle and LHCb experiments have measured numerous data of branching fractions and 
${\it CP}$ asymmetries of $B\to PP, PV$ decays, where $P(V)$ denotes a light
pseudoscalar (vector) meson. 
On the theoretical side, it requires complicated study of non-perturbative strong QCD dynamics in the charmless
 $B$ decays, which not only involve tree topologies but also   more complicated penguin loop diagrams.

Based on the leading order power expansion of $\Lambda_{QCD}/\mathrm{m}_b$, 
the QCD factorization (QCDF)\cite{Beneke:2000ry}, the perturbative QCD (PQCD)\cite{Lu:2000em},
and the soft-collinear effective theory (SCET)\cite{Bauer:2000yr} have been developed 
to study the charmless $B$ decays.
However, some puzzles encountered at the leading power of $\Lambda_{QCD}/\mathrm{m}_b$ 
in these   factorization approaches, for example, (I) the predicted branching fractions for   color-suppressed tree-dominated decays
 $\bar B^0\to\pi^0\pi^0$, $\rho^0\pi^0$ are too small comparing with experimental data, that is the so-called $\pi \pi$ puzzle,
(II) some direct {\it CP} asymmetries of $B \to PP$, $PV$ decays are inconsistent with experiment in signs, 
such as $K \pi$ puzzle.
Although some soft and sub-leading power of $\Lambda_{QCD}/\mathrm{m}_b$ effects were taken into account 
in the QCDF \cite{Cheng:2009cn} and the PQCD \cite{Li:2005kt}, the $B \to \pi \pi$ puzzle was still left in
the conventional factorization theorem. 
Unlike these perturbative approaches, some model-independent approaches were
introduced to analyze the charmless $B$ decays, such as global $SU(3)/U(3)$ flavor symmetry analysis \cite{he} and
flavor topological diagram approach based on flavor $SU(3)$ symmetry\cite{Cheng:2014rfa}.
Nowadays, $SU(3)$ breaking effects have to be considered to compare the
theoretical results with the precise experimental data.
It is also observed in the flavor topological diagram analysis that they have to fit three different sets of parameters for
the three types of $B$ decays respectively \cite{Cheng:2014rfa}
due to large difference between pseudo-scalar and vector final states of $B\to PP$, $B\to PV$ and $B\to VP$ decays.
There are too many parameters to be fitted thus its prediction power is limited.

In view of the above complexity and incompleteness in power correction
of factorization approaches and the limitation of the conventional flavor
topological diagram approach, a new method called
factorization-assisted-topological-amplitude (FAT) approach was proposed
in studying the two-body hadronic decays of $D$ mesons \cite{Li:2012cfa,Li:2013xsa}.
Aiming  to include all non-factorizable QCD contributions
compared to factorization approaches, it adopts the formalism of flavor topological diagram approach.
However, different from the conventional
flavor topological diagram approach, it had included $SU(3)$ breaking effect
in each flavor topological diagram assisted by factorization hypothesis, 
further reducing the number of free parameters by fitting all the decay channels 
and the precision of the FAT approach then not limited to the order of flavor $SU(3)$ breaking effect. 
In the following, we will analyze the charmless $B \to PP$, $PV$ decays in the FAT approach.

\section{The Amplitudes of $B \to PP$, $PV$ decays in FAT Approach}\label{sec:2}

The charmless two body  $B$ decays are induced by the quark level diagrams classified by 
leading order (tree diagram) and 1-loop level (penguin diagram) weak interactions.
For different $B$ decay final states, the tree level weak decay diagram can contribute via different orientations: 
the so-called  color-favored tree emission diagram $T$,
color-suppressed tree emission diagram $C$,
$W$-exchange tree diagrams $E$ and
$W$-annihilation tree diagrams $A$,  respectively. 
Similarly, the 1-loop penguin diagram can also be classified as  5-types:
color-favored QCD penguin emission diagram $P$,
color-suppressed QCD penguin emission diagram $P_C$,
penguin-annihilation  diagram $P_A$,
the time-like penguin   diagram $P_E$ and
electro-weak penguin emission diagram $P_{EW}$.
The three categories of  
$B \to PP$, $PV$ and $VP$ decays parameterized as three sets of parameters in the conventional topological diagram approach, will be parameterized as only one set of universal parameters  in the FAT approach. 

The $T$ topology is proved factorization to all orders of $\alpha_{s}$ expansion in QCD factorization approaches and SCET. 
Their numerical results also agree to each other in different approaches. Thus, to reduce one free parameter,  
we will just use their theoretical results from QCD calculation, not fitting from the experiments:
\begin{eqnarray}\label{eq:T}
T^{P_{1}P_{2}}&=&i\frac{G_{F}}{\sqrt{2}}V_{ub}V_{uq^{'}}a_{1}
(\mu)f_{p_{2}}(m_{B}^{2}-m_{p_{1}}^{2})F_{0}^{BP_{1}}(m_{p_{2}}^{2}),\nonumber\\
T^{PV}&=&\sqrt{2}G_{F}V_{ub}V_{uq^{'}}a_{1}
(\mu)f_{V} m_{V}F_{1}^{B-P}(m_{V}^{2})(\varepsilon^{*}_{V}\cdot p_{B}),\nonumber\\
T^{VP}&=&\sqrt{2}G_{F}V_{ub}V_{uq^{'}}a_{1}
(\mu)f_{P} m_{V}A_{0}^{B-V}(m_{P}^{2})(\varepsilon^{*}_{V}\cdot p_{B}),
\end{eqnarray}
where the superscript  of $T^{P_{1}P_{2}}$ denote the final mesons
with two pseudoscalar mesons, and $T^{PV(VP)}$ for recoiling mesons are
pseudoscalar meson (vector meson) with one pseudo-scalar and one vector meson final states.
 $a_1(\mu)$ is the effective Wilson coefficient of four quark operators with  QCD corrections.
 $f_{P_{2}}$($f_{P}$) and $f_{V}$ are the decay constants of the emitted pseudoscalar meson
and vector meson, respectively.   $F_{0}^{BP_{1}}$ ($F_{1}^{B-P}$) and $A_{0}^{B-V}$ are the
 form factors of $B\to P$ and $B\to V$ transitions, respectively.
$\varepsilon^{*}_{V}$ is the polarization vector of vector meson and
$p_{B}$ is the 4-momentum of $B$ meson.
For the color suppressed     $C$ topology, 
we parameterize its magnitude and associate phase as
$\chi^{C}$ and $\mathrm{e}^{i\phi^{C}}$ in $B\to PP$, $VP$ decays and
$\chi^{C^{\prime}}\mathrm{e}^{i\phi^{C^{\prime}}}$ in $B\to PV$, respectively
to distinguish cases in which the emitted meson is pseudo-scalar or vector meson:
\begin{eqnarray}\label{eq:C}
C^{P_{1}P_{2}}&=&i\frac{G_{F}}{\sqrt{2}}V_{ub}V_{uq^{'}}\chi^{C}\mathrm{e}^{i\phi^{C}}
        f_{p_{2}}(m_{B}^{2}-m_{p_{1}}^{2})F_{0}^{BP_{1}}(m_{p_{2}}^{2}),\nonumber\\
C^{PV}&=&\sqrt{2}G_{F}V_{ub}V_{uq^{'}}\chi^{C^{\prime}}\mathrm{e}^{i\phi^{C^{\prime}}}
         f_{V} m_{V}F_{1}^{B-P}(m_{V}^{2})(\varepsilon^{*}_{V}\cdot p_{B}),
\nonumber\\
C^{VP}&=&\sqrt{2}G_{F}V_{ub}V_{uq^{'}}\chi^{C}\mathrm{e}^{i\phi^{C}}
        f_{P} m_{V}A_{0}^{B-V}(m_{P}^{2})(\varepsilon^{*}_{V}\cdot p_{B}),
\end{eqnarray}
where the decay constants and form factors $f_{P}$, $f_{V}$,$F_{0}^{BP_{1}}$, $F_{1}^{B-P}$ 
and $A_{0}^{B-V}$ characterizing the $SU(3)$ breaking effects are factorized out.
The W-exchange $E$ topology is non-factorizable in QCD factorization approach that is expected smaller
than emission diagrams as power suppressed. We use $\chi^{E}$, $\mathrm{e}^{i\phi^{E}}$ to represent
the magnitude and its strong phase for all decay modes:
\begin{eqnarray}\label{eq:E}
E^{P_{1}P_{2}} &=&i\frac{G_{F}}{\sqrt{2}}V_{ub}V_{uq^{'}} \chi^{E} \mathrm{e}^{i\phi^{E}}
f_{B}m_{B}^{2}(\frac{f_{p_{1}}f_{p_{2}}}{f_{\pi}^{2}}),\nonumber\\
E^{PV,VP} &=&\sqrt{2}G_{F}V_{ub}V_{uq^{'}}\chi^{E} \mathrm{e}^{i\phi^{E}}
(\mu)f_{B}m_{V}(\frac{f_{P}f_{V}}{f_{\pi}^{2}})(\varepsilon^{*}_{V}\cdot p_{B}).
\end{eqnarray}
We will ignore $A$ topology, as its contribution is negligible as discussed in \cite{Cheng:2014rfa}.

Similarly, we parameterize the corresponding penguin diagrams with 8 parameters: 
chiral enhanced penguin amplitude $\chi^P$ and its phase $\phi^P$ excluding 
the factorizable leading power contribution of the $P$ topology,
flavor singlet penguin amplitude $\chi^{P_C}$, $\chi^{P_C^{\prime}}$ and 
their phases $\phi^{P_C}$, $\phi^{P_C^{\prime}}$ for the pseudo-scalar and vector meson emission, respectively,
the penguin annihilation amplitude $\chi^{P_A}$ and its phase $\phi^{P_A}$ for the vector meson emission only. 
The contribution from $P_E$ diagram is argued smaller than $P_A$ diagram, which
can be ignored reliably in decay modes not dominated by it. 
Similar to $T$ and leading power contribution from the $P$ topology, 
we calculate $P_{EW}$ topology,  the largest contribution from EW-penguin contribution, in QCD factorization approaches. 
%Totally, 14 parameters induced in the amplitudes of tree and penguin topological amplitudes of 
%$B \to PP$ and $B \to PV$ decays will be fitted together.

\section{Numerical results and discussion}\label{sec:3}

With the experimental data of 37 branching fractions and 11 ${\it CP}$ asymmetry parameters\cite{PDG}, 
we do a global fit to extract the 14 parameters.  The best-fitted values and the corresponding $1\sigma$ uncertainty are:
\begin{equation} \label{parameter}
\begin{array}{cccc}
 \chi^{C}=0.48 \pm 0.06,&\phi^{C}=-1.58 \pm 0.08,&
\chi^{C^{\prime}}=0.42 \pm 0.16,&\phi^{C^{\prime}}=1.59\pm 0.17\\ 
\chi^{E}=0.057\pm 0.005,& \phi^{E}=2.71\pm 0.13,&
\chi^{P}=0.10\pm 0.02,&\phi^{P}=-0.61\pm 0.02,\\ 
 \chi^{P_C}=0.048 \pm 0.003,& \phi^{P_C}=1.56 \pm 0.08,&
 \chi^{P_C^{\prime}}=0.039\pm 0.003,& \phi^{P_C^{\prime}}=0.68 \pm 0.08,\\
 \chi^{P_A}=0.0059\pm 0.0008,& \phi^{P_A}=1.51\pm 0.09,&&
 \end{array} 
\end{equation}
with $\chi^{2}/d.o.f=45.2/34=1.3$.   
This $\chi^{2}$ per degree of freedom is smaller than the conventional flavor diagram approach \cite{Cheng:2014rfa},
even though with much more parameters than us.
The mapping of well-known QCDF-amplitudes introduced in \cite{Beneke:2003zv} and topological diagrams
 amplitudes in FAT approaches were compared  in table \ref{FATQCDF}.
It is apparent that there are large differences between results fitted from experimental data 
 in the FAT approaches and the calculated results in the QCDF, especially for the strong phases. 
Later we will show that the small strong phases, $\phi^{C}$ and $\phi^{C^{\prime}}$  
from QCDF are the main reason for the $\pi \pi$ and $\pi K$ puzzles.

\begin{table}[tbp]
\caption[]{The amplitudes and strong phases of topological diagrams in the FAT corresponding to contributions in the QCDF.}
  \vspace{0.4cm}
  \setlength\tabcolsep{2.5pt}
\begin{center}
\label{FATQCDF}
\footnotesize\begin{tabular}{|c||c|c|c|c|c|c|c|c|c|}
\hline
Diagram  & T  &  C  & $P_C$  & P(PP)
& $P_{EW}$ &  E  &  A  & $P_A$(PV)  &$P_E$  \\
\hline
FAT   & $a_1$  & $\chi^{C^{(\prime)}}\mathrm{e}^{i\phi^{C^{(\prime)}}}$ & $\chi^{P_C^{(\prime)}}\mathrm{e}^{i\phi^{P_C^{(\prime)}}}$ &$a_{4}(\mu)+ \chi^{P}\mathrm{e}^{i\phi^{P}}r_{\chi}$
 &  $a_9(\mu)$ & $\chi^{E} \mathrm{e}^{i\phi^{E}} $  & $-$  & $-i \chi^{P_A}\mathrm{e}^{i\phi^{P_A}}$ &- \\
% \hline
 &-&$0.48\mathrm{e}^{-1.58i}$  &$0.048\mathrm{e}^{1.56i}$&$-0.12\mathrm{e}^{-0.24i}$&-0.009&$0.057\mathrm{e}^{2.71i}$&&
 $0.0059\mathrm{e}^{-0.006i}$&\\
 \hline
QCDF & $\alpha_1$ &$\alpha_2$ &$\alpha_3$ & $\alpha_4$ & $\alpha_3^{\mathrm{EW}}$ &$\beta_1$ &$\beta_2$ &$\beta_3$& $\beta_4$\\
%\hline
&-&$0.22\mathrm{e}^{-0.53i}$  &$0.011\mathrm{e}^{2.23i}$&$-0.089\mathrm{e}^{0.11i}$&$-0.009 \mathrm{e}^{0.04i}$&0.025&-0.011
&-0.008&-0.003\\
 \hline
\end{tabular}
\end{center}
\end{table}

Using the fitted parameters in eq.(\ref{parameter}), we give the numerical results of branching fractions 
and the direct ${\it CP}$ and mixing-induced ${\it CP}$ asymmetries of charmless $B_{(s)}\to PP, PV$ decays shown in the tables of ref.\cite{Zhou:2016jkv} 
Nearly 100 channels are   provided to be tested in the future experiments.
Similar to the conventional topological diagram approach \cite{Cheng:2014rfa},
the long-standing puzzle of large $B^0\to \pi^{0}\pi^{0}$  branching ratio can be resolved well attributed to
the appropriate magnitude and phase of $C$ in FAT approach compared with the small magnitude of 
$C=\alpha_2=0.20^{+0.17}_{-0.11}$ by perturbative calculation in QCDF. 
However, $|T^{\pi \pi}|:|C^{\pi \pi}|=1:0.47$ in FAT approach is not as large as  the one in ref.\cite{Cheng:2014rfa},  
where the ratio even reached 0.97 in Scheme C.
The branching fractions of pure penguin decays  $B^{-}\to K^{-}K^{0}$, $B^{0}\to K^{0}\bar{K^{0}}$ 
given in the FAT approach are in much better agreement with experimental data than 
the previous conventional flavor diagram approach \cite{Cheng:2014rfa}, 
as we have considered the flavor $SU(3)$ breaking effect. 
With a large strong phase for sub-leading contribution $C$ in FAT approach, 
the $K \pi $ puzzle can also resolved.
This again implies large power corrections or large non-perturbative QCD corrections 
in the $C$ diagram of $B\to \pi K$ decays.

The flavor $SU(3)$ breaking effect considered here in every topology
amplitude between $B \to \pi\pi $ and $B \to \pi K $ is around $10\%$  
and larger than $20\%$ in corresponding $B \to PV$ models.
The difference between $\pi$ and $\rho$ meson emission is indeed much larger than the so called flavor $SU(3)$ breaking effect between $\pi$ and $K$ meson due to the meson decay constant $f_\rho > f_K$ and 
more larger characterized by the $ K $ and $ K^*$  decay constant.
\section{Conclusions}\label{sec:5}

We studied charmless two-body hadronic $B$ decays in factorization assisted
topological amplitude approach. By using the factorization results for $T$
and $P_{EW}$ diagrams,
there were 6 parameters $\chi^{C}(\phi^{C}),\chi^{C^{\prime}}(\phi^{C^{\prime}})$
and $\chi^{E}(\phi^{E})$ for tree diagrams $C,E$ and 8 parameters $\chi^{P}(\phi^{P}),
\chi^{P_C}(\phi^{P_C}), \chi^{P_C^{\prime}}(\phi^{P_C^{\prime}})$ and $\chi^{P_A}(\phi^{P_A})$
for QCD-penguin diagrams to be fitted from 48 measured data of branching ratios and {\it CP} asymmetry parameters
of the $B \to PP$, $PV$ decays together. The $\chi^{2}$ per degree of freedom is smaller than the conventional flavor diagram approach,  even with much more free parameters in their approach.  
With the fitted parameters, we predicted branching fractions of  nearly 100 charmless $B_{(s)} \to PP$,  $PV$
decay modes and their $CP$ asymmetry parameters. 
The long-standing puzzles of $\pi\pi$ branching ratios and $\pi K$ $CP$ asymmetry have been resolved consistently
with not too large color suppressed tree diagram contribution $\chi^{C}$. 
The flavor $SU(3)$ breaking effect
between $\pi$ and $K$ were approximately $10\%$, even more than $20\%$
in $\rho$ and $K^{*}$ meson case.

\section*{Acknowledgments}
%We are grateful to Hsiang-nan Li, Wei Wang, Rui Zhou, Fusheng Yu, Ying Li and Qin Qin for useful discussion.
%We also thanks Fred James providing help for analyzing error bar in Minuit program.
The work is partly supported by National   Science Foundation of China (11375208, 11521505, 11621131001 and 11235005).

\end{document}